\documentclass[12pt]{article}

\usepackage{amsmath}
\usepackage{amsfonts}
\usepackage{amssymb}
\usepackage{amscd}

\newcommand{\refeq}[1]{(\ref{#1})}
\newcommand{\rmd}{\mathrm{d}}
\newcommand{\Lieg}{\mathfrak{g}}
\newcommand{\G}{\mathcal{G}}
\newcommand{\cX}{\mathcal{X}}
\newcommand{\id}{\mathrm{id}}

\newcommand{\cf}{{\it cf.}}
\newcommand{\eg}{{\it e.g.}}
\newcommand{\ie}{{\it i.e.}}

\newcommand{\tE}{\widetilde{E}}
\newcommand{\tM}{\widetilde{M}}
\newcommand{\trho}{\widetilde{\rho}}
\newcommand{\ttE}{{\widetilde{\widetilde{E}}}}

\newcommand{\ttrho}{\widetilde{\widetilde{\rho}}}
\newcommand{\tp}{\widetilde{p}}
\newcommand{\tpi}{\widetilde{\pi}}

\newcommand{\tGamma}{\widetilde{\Gamma}}
\newcommand{\te}{\widetilde{e}}
\newcommand{\tte}{{\widetilde{\widetilde{e}}}}
\newcommand{\tpsi}{\widetilde{\psi}}
\newcommand{\ttpsi}{\widetilde{\widetilde{\psi}}}
\newcommand{\s}{\sigma}

\begin{document}

\title{Lie Algebroid Yang Mills 
with Matter Fields} 
\author{C.~Mayer and T.~Strobl} 
\date{\today}
\maketitle

\begin{abstract}
Lie algebroid Yang-Mills theories are a generalization of Yang-Mills
gauge theories, replacing the structural Lie algebra by a Lie
algebroid $E$.  In this note we relax the conditions on the fiber
metric of $E$ for gauge invariance of the action functional. Coupling
to scalar fields requires possibly nonlinear representations of Lie
algebroids. In all cases, gauge invariance is seen to lead to a
condition of covariant constancy on the respective fiber metric in
question with respect to an appropriate Lie algebroid connection.

The presentation is kept in part explicit so as to be accessible also
to a less mathematically oriented audience.
\end{abstract}

\tableofcontents

\section{Introduction}
In Ref.~\cite{Strobl} pure Yang Mills (YM) gauge theories have been
generalized to a setting where the structural Lie algebra is replaced
by a Lie algebroid. This is a vector bundle $E\to M$ with, among
others, a Lie algebra structure on its sections, thus reducing to a
Lie algebra for $M$ being a point, in which case also the Lie
Algebroid Yang Mills (LAYM) gauge theory reproduces just an ordinary
YM theory in $d$ spacetime dimensions. Simultaneously, it constitutes
a nonlinear type of gauge theory, which in contrast to topological
prototypes like the Poisson Sigma Model \cite{Schaller:1994es,
  Ikeda:1993fh} has propagating degrees of freedom. At least on the
classical level, moreover, these propagating degrees seem to be those
of ordinary YM theories, albeit of potentially different type and with
potentially different structure groups, glued together over some
finite dimensional moduli space \cite{Strobl}.

In this paper we first reconsider these LAYM theories, using a second
type of gauge symmetries (one that is induced by an auxiliary
connection chosen on $E$). For actions of type $F_{(1)}^2 +
F_{(2)}^2$, where $F_{(1)}$ and $F_{(2)}$ denote the 1-form and 2-form
field strengths of the gauge field, respectively, and the square is
understood as denoting an appropriate norm square, we find that gauge
invariance restricts $E$ to be an action Lie Algebroid, which from the
physical point of view corresponds to ordinary YM theory coupled to
``Higgs fields'' possibly taking values in some curved target
manifold. We then show that the action \cite{Strobl} of the form $B \,
F_{(1)} + F_{(2)}^2$, $B$ denoting Lagrange multiplier fields, is
gauge invariant if the respective fiber metric on $E$ is covariantly
constant w.r.t.~a Lie algebroid or $E$-connection (induced by the
auxiliary ordinary one on $E$) that (at least when flat) 
can be thought of generalizing the adjoint representation of a Lie algebra.

Then we turn to the main subject of the paper, the coupling of scalar
matter fields to the LAYM theory. We first assume that these scalar
fields take values in some vector bundle $V \to M$. Starting with some
elementary ansatz for gauge transformations of the scalar fields, we
are lead rather directly to flat $E$-connections on $V$. This is the
mathematical generalization of a Lie algebra representation on a
vector space, to which it reduces for $M$ being a point. Gauge
invariance of the kinetic term requires that the fiber metric on $V$,
needed for its construction, is covariantly constant w.r.t.~this
$E$-connection.

This is then, in a second step, generalized to scalar fields with
target space an arbitrary bundle $p \colon \tilde M \to M$ over the
base of the Lie algebroid. In this context it is helpful to observe
that all the needed data can be reassambled into a Lie algebroid
structure on $\tE = p^*E$. It is the generalization of an action Lie
algebroid to the context of Lie algebroids. Gauge invariance of the
more general kinetic term in this sigma model context now requires a
fiber metric on $V\tM$ (the bundle of vertical vectors on $\tM$) that
is covariantly constant w.r.t.~a canonically induced flat
$\tE$-connection on $V\tM$.

This perspective suggests a reinterpretation of the scalar
fields. Namely, one may have started right away by considering the Lie
algebroid $\tE \to \tM$. The previously constructed coupled
LAYM-matter Lagrangian is then seen as a functional of a (pure, if one
likes) LAYM gauge theory for $\tE$ of the form that only \emph{part}
of the 1-form field strengths enter the functional with Lagrange
multipliers while the ``remaining'' ones\footnote{This is formulated
more intrinsically in the last section of the present article.} are
squared by means of an appropriate symmetric covariant two-tensor on
$\tM$ (a partially degenerate ``metric'' tensor on $\tM$).

Finally, we exploit generalized Bianchi identities to further relax
the condition on the fiber metric ${}^E g$ on $E$ in a pure
LAYM-theory of the type considered in \cite{Strobl}. In fact, for
gauge invariance it turns out to be sufficient that the restriction of
${}^E g$ to the kernel of the anchor map of $E$ is invariant w.r.t.~a
canonical Bott-type $E$-connection.

\section{Lie Algebroid Yang Mills revisted}

Here we start recalling the basic elements of a Lie algebroid
Yang-Mills (YM) theory in a rather explicit, elementary fashion. The
$d$-dimensional spacetime manifold we denote as $(\Sigma,h)$, where $h$
is a fixed (possibly pseudo-) Riemannian metric. The structural Lie
algebra entering the construction of an ordinary YM algebra is
generalized to a Lie algebroid $(E\rightarrow M, [ \cdot,\cdot],
\rho)$, the basic definition of which (together with other background
material) can be found in Appendix~A. Using some local coordinates
$x^i$ on $M$ and a local frame $e_a$ of $E$, all the structural
quantities of $E$ can be described by functions $\rho_a^i(x)$ and
$C_{ab}^c(x)$, satisfying the differential equations
\begin{equation} \label{Jakobi}
  \rho_a^j\rho_{b,j}^i - \rho_b^j\rho_{a,j}^i = C_{ab}^c\rho_c^i\;,\qquad
  C_{ad}^eC_{bc}^d + \rho_a^iC_{bc}^e{}_{,i} + cycl(abc)=0\;.
\end{equation} 
Clearly, if $M$ is a point, thus $C_{ac}^c$ not depending on $x^i$, and
$\rho_a^i\equiv 0$, one reobtains the structure constants of a Lie
algebra $\Lieg$.

For the fields we take 1-form fields $A^a = A^a_\mu(u^\mu) \rmd u^\mu$,
where $u^\mu$ are coordinates on $\Sigma$, together with 0-form fields
$X^i(u^\mu)$. The latter ones describe a map $\cX^i$ from $\Sigma$ to $M$, $X$ and
$A$ together a vector bundle map  $a \colon T\Sigma \to E$ (\cf~\cite{BKS}
or \cite{Strobl} for further details).  Associated to these ``gauge
fields'' are the ``field strengths''
\begin{equation}
  \label{eq:f}
  \begin{split}
    F^i &= \rmd X^i - \rho^i_aA^a\\
    F^a &= \rmd A^a + \tfrac{1}{2}\,c_{bc}^a A^b\wedge A^c + 
    \Gamma_i{}_b^aF^i\wedge A^b \, ,
  \end{split}
\end{equation}
where $\Gamma_i{}_b^a$ are the coefficients of a fixed background
connection $\nabla_i$ in $E$; they are necessary if one wants to
define the 2-form field strengths $F^a$ covariantly with respect to
$E$-frame changes.\footnote{Both field strengths together can be given
a meaning also without introducing a connection 
(\cf,~\eg~\cite{char}); it is only the
separation of the 2-form part which requires the connection.---The
fixed connection on $E$ is not to be confused with the gauge fields,
which, in the case of an ordinary YM theory are connections in a
principal bundle; the former ones correspond to structures needed to
be fixed for defining a functional, while the latter ones are
dynamical, \ie~they are the argument of that functional.} This
becomes most transparent when rewriting the second equation according
to 
\begin{equation}
 F^a = (D_\Gamma A)^a -
 \tfrac{1}{2}\,T_{bc}^aA^b\wedge A^c \, , \label{eq:F}
\end{equation}
where 
\begin{eqnarray}
  (D_\Gamma A)^a &\equiv& \rmd A^a +  \Gamma_i{}_b^a \,\rmd X^i \wedge A^b \; , \\
  \label{eq:T1}
  T_{ab}^c &\equiv& -C_{ab}^c + \rho_a^i\Gamma_i{}_b^c - \rho_b^i\Gamma_i{}_a^c\;.
\end{eqnarray}
Here $D_\Gamma A$ is the exterior covariant derivative on $A \in
\Omega(\Sigma, \cX^*E)$ and $T$ is the $E$-torsion of the $E$-connection
$\nabla_{\! \rho(\cdot)}$, both being induced by the chosen connection
$\nabla$ on $E$ (\cf~Appendix~A for further details); the 2-form field
strength is then an element in $\Omega^2(\Sigma, \cX^*E)$. In the
specific case described at the end of the previous paragraph, 
$E$~$=$~$\Lieg$, one is back to the usual YM setting (with a Lie algebra
valued 2-form curvature and no 1-form field strength)\footnote{We discuss
trivial bundels over $\Sigma$ here only, \cf~\cite{char} for how to
generalize to nontrivial ones.}. This also applies to the gauge
transformations, which we will now address.

Infinitesimally the gauge transformations are taken to be of the form
\begin{eqnarray}
  \delta_\epsilon X^i &=& \rho^i_a \epsilon^a \label{eq:gt1}\\
    \delta_\epsilon A^a &=& \rmd \epsilon^a + C^a_{bc}A^b\epsilon^c + 
    \Gamma_{ib}^a\epsilon^b F^i \, , \label{eq:gt2}
  \end{eqnarray}
where the same connection coefficients were used that entered
already the definition of $F^a$. There is also an alternative, geometrically
motivated, off-shell closed version of gauge symmetries, not using an
auxiliary connection and also generalizing the usual YM ones
(\cf~\cite{BKS,Strobl}); as mentioned already in the Introduction, in
this note we want instead to focus on this connection-induced 
type of gauge symmetries. In any
case, in the variation of $A^a$ a term proportional to $F^i$ is needed
for $E$-covariance again. Note, however, that the terms in
\refeq{eq:gt2} do not combine completely into covariant objects
following the pattern of
\refeq{eq:F}: 
\begin{equation}
 \delta_\epsilon A^a = D_\Gamma  \epsilon^a - T^a_{bc}A^b\epsilon^c - 
\rho_c^i \epsilon^c\Gamma_i{}_b^a A^b \, .
\end{equation}
The reason is that infinitesimal gauge transformations are a
derivative-type object and the extra term is needed for compatibility
with \refeq{eq:gt1}, \cf~\cite{BKS} as well as the likewise discussion
following Eq.~\refeq{eq:5} below.

On the $A$-fields the variations (\ref{eq:gt2}) close only modulo a term 
proportional to $F^i$,\footnote{These equations hold in a frame where the
parameters $\epsilon^a$ depend on coordinates of $\Sigma$ only, but not also
on the fields $X$ or even $X$ and $A$. We intend to provide a more coordinate 
independent interpretation elsewhere.}
\begin{eqnarray}
  \label{var}
  \Big(\big[\delta_{\epsilon_1}, \delta_{\epsilon_2}\big] -
  \delta_{\epsilon_3}\Big)A^a
  &=& 
  \epsilon_1^b\epsilon_2^cF^i S_{i}{}_{bc}^a\,,\\
\epsilon_3^a &\equiv& C_{bc}^a\epsilon_1^b\epsilon_2^c \label{epsilon3}\\
  \label{S}
  S_i{}_{bc}^a 
  &\equiv&
  \nabla_iT_{bc}^a + \rho_c^jR_{ij}{}_b^a -
  \rho_b^jR_{ij}{}_c^a
\end{eqnarray}
where $\nabla_i$ denotes the covariant derivative with respect to the
fixed background connection on $E$ and $R_{ij}{}_a^b$ its
curvature. As a consequence, the gauge
symmetries provide a representation of the Lie algebroid on the fields
$X^i$, $A^a$ only if either $F^i=0$ or $S_i{}_{bc}^a=0$.  For later
use we provide the gauge variation of the field strengths:
\begin{eqnarray}
  \label{eq:4a}
    \delta_\epsilon F^i &=& \epsilon^a\big(\nabla_j \rho_a^i \big)F^j\;,\\
    \delta_\epsilon F^a &=& -\epsilon^c\big(c^a_{bc} 
    + \Gamma_i{}_c^a\rho_b^i\big)F^b\,, \nonumber\\ \label{eq:4b}
    && +\, \tfrac{1}{2}\,\epsilon^bR_{ij}{}_b^aF^i\wedge F^j
    + \epsilon^c S_i{}_{bc}^a F^i\wedge A^b\,.
\end{eqnarray}
where $\nabla_j \rho_a^i\equiv \rho_{a,j}^i-\Gamma_{ja}^b\rho_b^i$ denotes the covariant derivative w.r.t.~the index $a$ only.

With these ingredients it is now easy to provide a generalization of
BF-theories to the setting of Lie algebroids
(\cf~\cite{Grav,Strobl,Bonechi-Zabzine}):
\begin{equation} \label{LABF}
  S_{LABF} = \int_\Sigma B_i\wedge F^i + B_a\wedge F^a\,,
\end{equation}
where $B_i$ and $B_a$ are $(d-1)$ and $(d-2)$-form fields,
respectively, the transformations of which can be adjusted to render
the action invariant under the above gauge transformations. The field
equations $F^i=0$ and $F^a=0$ require $(X,A)$ to correspond to a Lie
algebroid morphism from $T\Sigma$ to $E$, while the above gauge
transformations reduce to Lie algebroid homotopies in that case
(\cf~\cite{BKS} for further details).

So as to construct a gauge invariant Lie Algebroid YM action, one would
naturally be lead to square both field strengths,
\begin{equation}
  \label{Fhoch2}
  \int_\Sigma -\tfrac{1}{2}\; F^a\wedge\star F^bg_{ab} 
  - \tfrac{1}{2}\; F^i\wedge\star F^j g_{ij}\;,
\end{equation}
using a metric $g \sim g_{ij}$ on $M$, a fibre metric $^Eg \sim
g_{ab}$ on $E$, and the metric $h$ on $\Sigma$ for the Hodge dual of
differential forms. The condition of gauge invariance of the action
should then imply some meaningful conditions on the additional
structures $g_{ij}$, $g_{ab}$ (generalizing ad-invariance of the
metric on the Lie algera in the ordinary YM case) and their existence
then possibly a restriction on the possible Lie algebroids $E$
(quadratic Lie algebras in the YM situation).

In the context of the above functional, however, the restrictions turn
out to be enormous, bringing one back implicitely to the realm of
ordinary YM gauge theories: The variation of the field stregth $F^a$ in
the first term produces terms proportional to $F^i\wedge
F^j\wedge\star F^a$ and $F^i\wedge A^a\wedge\star F^b$, both of which
cannot be compensated for by variations of other parts of the actions
and thus have to vanish individually. The vanishing of the first term
implies that $\nabla$ is a flat connection on $E$, $R_{ij}{}_a^b=0$,
the second constraint, $S_i{}_{bc}^a=0$, then reduces to
$\partial_iC_{bc}^a=0$ (\cf~Eqs.~(\ref{S}) and (\ref{eq:T1})). This in
turn implies that we can identify $E$ with $M \times\Lieg$, $\Lieg$
being the Lie algebra with the respective structure constants
$C_{bc}^a$ and $\rho \colon E \to TM$ can be identified with a
representation of it on $M$ (\cf~Eq.~(\ref{Jakobi})).\footnote{Such an
  $E$ is called an action Lie algebroid.} From a physical perspective,
then, the theory reduces to standard YM theory (first term in
(\ref{Fhoch2})) with structural Lie algebra $\Lieg$, coupled to a
Higgs-type sigma model with the Higgs fields taking values in $M$ (the
second term in (\ref{Fhoch2}) reduces to the usual kinetic term of
such a theory).

In fact, part of these conditions, namely $S_i{}_{bc}^a=0$, can
already be deduced from (\ref{var}), taking into account that
obviously $F^i=0$ are not field equations for the action functional
(\ref{Fhoch2}). This consideration, however, provides also a hint for
a way to avoid the above no-go-type result: One may want to ensure
that $F^i=0$ \emph{are} part of the field equations of the strived for
generalization of the YM-action (note that for an ordinary YM theory,
$M$ is a point and $F^i$ vanishes identically). In this way one is
lead to \cite{Strobl} \begin{equation}
  \label{eq:LAYM}
  S_{LAYM} = \int_\Sigma B_i\wedge F^i 
  - \tfrac{1}{2}\; F^a\wedge\star F^bg_{ab}\;,
\end{equation}
$B_i$ being $(d-1)$ forms on $\Sigma$ like in (\ref{LABF}) above.

Now again we ask for the conditions on the structural ingredients,
\ie~$E$, $\nabla_i$, and $g_{ab}$, such that the above functional is
gauge invariant w.r.t.~the symmetries generated by Eqs.~(\ref{eq:gt1},
\ref{eq:gt2}) (for some transformation induced on the
$B_i$-fields). The action functional (\ref{eq:LAYM}) is gauge
invariant w.r.t.~those gauge transformations, if the fiber metric
$^Eg \sim g_{ab}$ is covariantly constant w.r.t.~a certain Lie
algebroid (``$E$-'') connection $^E\Tilde{\nabla}$,\footnote{This geometric 
interpretation was
observed already shortly after completion of \cite{Strobl} and
reported \eg~in~\cite{Peruggia}.}
\begin{equation} {}^E\Tilde{\nabla} \, {}^E\!g = 0 \, .
\label{condition}
\end{equation}
This $E$-connection is one induced by the ordinary connection $\nabla$
on $E$ and defined via\footnote{The concept of a Lie algebroid connection and
corresponding generalizations of curvature and torsion is recalled in
Appendix A.}
\begin{equation}  \label{tildecon} 
{}^E\Tilde{\nabla}_\psi \Tilde{\psi} = 
\nabla_{\!\rho(\widetilde{\psi})} \psi + [\psi, \Tilde{\psi}] \, .
\end{equation}
In local components the coefficients of this $E$-connection read as
$\tGamma^a{}_{bc}=\rho_c^i\Gamma_{ib}^a+c_{bc}^a
=\rho_b^i\Gamma_{ic}^a-T_{bc}^a$.
 From the first equality one obtains
\refeq{condition} at once, observing that the first line of
\refeq{eq:4b} contains precisely $\tGamma^a{}_{bc}$ (while
the two terms in the second line, which resulted in the unwanted
severe restriction on $E$ in the case of \refeq{Fhoch2}, now can be
absorbed by the variation of $B_i$ since they are both proportional to
$F^i$); the second equality shows that ${}^E\Tilde{\nabla}$ differs
from the more obvious $E$-connection $\nabla_{\rho(\cdot)}$ by
subtraction of its own $E$-torsion.

Note that for $M$ being a point, the first term in \refeq{tildecon} is
absent since $\rho$ vanishes and one reobtains the usual condition of
an ad-invariant metric on the Lie algebra. The \emph{existence} of an
ordinary connection $\nabla$ and a fiber metric $^Eg$ such that
(\ref{condition}) is fulfilled, poses a restriction on $E$. In the
case of integrable Lie algebroids and for $^Eg$ having definite
signature, this restriction is conjectured by Fernandes to precisely
give Lie algebroids $E$ coming from \emph{proper} Lie groupoids (a
notion coinciding with compactness in the Lie group case)
\cite{ESI-Fernandes}. It is amusing that this particular
$E$-connection pops out naturally from invariance of the functional
(\ref{eq:LAYM}) and the simple ansatz (\ref{eq:gt1},\ref{eq:gt2}) for
the gauge symmetries.

In fact, it turns out that a condition like \refeq{condition} (or the
likewise one found in \cite{Strobl}) is sufficient but not also
necessary for gauge invariance of the action functional $S_{LAYM}$. We
will discuss this issue in detail in section \ref{Discussion} below.

Before closing this section, we make a remark on some geometric 
interpretation of the tensor \refeq{S}; in fact it is related to 
the $E$-curvature of ${}^E\Tilde{\nabla}_a$ by contraction with 
the anchor map $\rho$ (\cf~Appendix~A): 
\begin{equation}\label{ER=rhoS}
{}^E\Tilde{R}_{ab}{}_c^d =
\rho_c^iS_i{}_{ab}^d \, . 
\end{equation} 
Hence, if the gauge transformatons close off-shell, \ie~if
$S_i{}_{bc}^a=0$, then ${}^E\Tilde{\nabla}_a$ is flat. The converse
statement is not true. We will encounter flat $E$-connections in the
subsequent section when considering the issue of coupling matter
fields to the above action functional $S_{LAYM}$. Flat $E$-connections
on vector bundles over $M$ are the natural generalization of a
(linear) representation of a Lie algebra to the context of Lie
algebroids (\cf,~\eg,~\cite{LecturesCrainicFernandes}).\footnote{We
will in the following section, however, \emph{not} assume familiarity
with such a mathematical concept. Instead, we will start in a
pedastrian style for the construction of a coupling to matter fields
and be lead automatically to the mathematical concepts by means of
gauge invariance.} 
A flat $E$-connection ${}^E\Tilde{\nabla}$ on
$E$ can then be considered as a possible generalization of the adjoint
representation of a Lie algebra.

\section{Matter Fields with values in vector bundels}

In this section we address the issue of coupling scalar fields to the
YM-type theory of the previous section. Since we address trivial
bundles over $\Sigma$ only within this note, in the ordinary YM
situation this would correspond to some functions on $\Sigma$ taking
values in a vector space which carries a representation of the
structural Lie algebra. Representations of Lie algebroids are known in
the mathematical literature as flat $E$-connections. Here we will,
however, adopt a more pedastrian, physics oriented route which will
lead us there by itself. In fact, following the same route we will be
lead to a more general setting, permitting also non-linear
representations.

For this purpose we start with a set $\phi^\sigma$ of functions on 
$\Sigma$. We expect/want formulas to be covariant w.r.t.
\begin{equation} \label{cov}
 \phi^\sigma \to \overline{\phi^\sigma} \equiv M^\sigma_\tau \phi^\tau
\end{equation}
for arbitrary matrices $M^\sigma_\tau$. In the usual YM setting this
corresponds to a change of basis in the representation space of the
Lie algebra. In the present more general setting the gauge fields
contain not only 1-forms $A$ on $\Sigma$, but also 0-forms $X^i$ and
it is thus natural to permit $M^\sigma_\tau$ to depend on $x$. More
abstractly, this implies that the Higgs-type scalar fields $\phi^\sigma$
correspond to sections of $\cX^*V$, where $V$ is a vector bundle over
$M$, the same base as the Lie algebroid $E$ (and $\cX$ the previous map
from $\Sigma$ to $M$).


Now we make the following ansatz for infinitesimal gauge
transformations:
\begin{equation}
  \label{eq:5}
  \delta_\epsilon\phi^\sigma = -\epsilon^a\Gamma_a{}^\sigma_\tau\phi^\tau\;,
\end{equation}
where $\Gamma_a{}^\sigma_\tau$ are some at this point not further
specified fixed parameters depending on $X$. Covariance restricts them
further, however: We want that for $\widetilde{\phi^\sigma}$ we have a
likewise formula. On the other hand, using Eq.~\refeq{eq:gt1} and the
fact that $\delta_\epsilon(M^\sigma_\tau\phi^\tau)=
\delta_\epsilon(M^\sigma_\tau)\phi^\tau+M^\sigma_\tau \delta_\epsilon\phi^\tau$, we can
determine the transformation property of the above coefficients,
\begin{equation}
  \overline{\Gamma_a{}^\sigma_\tau}
  = 
  M^\sigma_{\sigma'}\,\Gamma_a{}^{\sigma'}_{\tau'} \,M^{-1}{}^{\tau'}_\tau - 
  \rho_a^i\,M^\sigma_{\sigma',i}\,M^{-1}{}^{\sigma'}_\tau\;.
\end{equation}
This implies that these coefficients have the geometrical
interpreation of an $E$-connection ${}^E\nabla$ on the vector bundle
$V$.\footnote{We usually drop the extra upper $E$ in the
  $E$-connection coefficients, since their indices already make clear
  of what nature they are. Solely with the respective $E$-curvatures
  we keep it for clarity also in the components. Note that in the
  present section ordinary connections as well as $E$-connections
  always refer to the vector bundle $V\to M$, in contrast to the
  previous section where they both referred to $E\to M$ itself---for
  notational simplicity we use the same symbols. The representation
  space $V$ can be chosen as $E$ itself, certainly; the notations are
  chosen such that they coincide in that particular case.}

Finally we demand that the gauge transformations close on the newly 
introduced fields, 
\begin{equation} \label{closure}
[\delta_{\epsilon_1},\delta_{\epsilon_2}]\phi^\sigma =
\delta_{\epsilon_3}\phi^\sigma\,,
\end{equation}
where $\epsilon_3$ is given by formula \refeq{epsilon3}. Note that in
this case it is not natural to permit a contribution proportional to
$F^i$ as in \refeq{var}, although using the metric $h$ on $\Sigma$ one
could produce also 0-form contributions from $F^i$. The
condition~\refeq{closure} is \emph{equivalent} to ${}^E\nabla$ having
vanishing $E$-curvature. Thus with these physical considerations we
indeed find that to couple matter fields to a Lie algebroid Yang-Mills
theory $S_{LAYM}$ for some given structural Lie algebroid $E \to M$ we
need a vector bundle $V \to M$ carrying a flat $E$-connection
${}^E\nabla$, a ``Lie algebroid representation'' on $V$ in the
mathematical sense. Note that vanishing $E$-curvature by definition
means $[{}^E\nabla_\psi , {}^E\nabla_{\widetilde{\psi}} ] =
{}^E\nabla_{[\psi,\widetilde{\psi}]}$, with the Lie algebroid bracket
on the r.h.s.; thus this indeed implies that the differential
operators ${}^E\nabla_\psi$ are a representation of the Lie algebra
defined by the Lie algebroid bracket.

The generalization of a covariant derivative on Higgs fields in
ordinary YM-theory takes the form \begin{equation}
  \label{eq:6}
  D\phi^\sigma = 
  \rmd\phi^\sigma + \Gamma_a{}_\tau^\sigma A^a\phi^\tau +
  \Gamma_i{}_\tau^\sigma \phi^\tau F^i\;,
\end{equation}
where $\Gamma_i{}_\tau^\sigma$ and $\Gamma_a{}_\tau^\sigma$ are
coefficients of an ordinary connection $\nabla$ and the above
$E$-connection ${}^E\nabla$, respectively, both defined on $V$ (it is
certainly their pullback by $\cX$ that enters in such an expression,
$\phi$ being a section in $\cX^*V$---following physics conventions such
identifications are understood). The first two terms are familiar ones
if, following Eq.~\refeq{eq:5}, one identifies $\Gamma_a{}_\tau^\sigma$ with
the coefficients of a representation; the contribution proportional to
$F^i$ is again needed for covariance (under changes of $E$- and
$V$-frames). Indeed, the terms in \refeq{eq:6} may be recombined into
\begin{equation}
  \label{eq:cov}
  (D\phi)^\sigma = (D_\Gamma \phi)^\sigma  - A^aT_a{}_\tau^\sigma\phi^\tau
\end{equation}
where $(D_\Gamma \phi)^\sigma = \rmd \phi^\sigma + \rmd
X^i\Gamma_i{}_\tau^\sigma\phi^\tau$ is the canonical exterior
covariant derivative in $\cX^*V$ induced by the connection $\nabla$ on
$V$. $T$, on the other hand, is (the pullback by $\cX$ of) a section in
$E^* \otimes End(V)$, defined, for any $\psi \in \Gamma(E)$, by means
of the difference of two $E$-connections (on $V$), namely
\begin{equation}
  \label{eq:8}
  T_{\psi} = \nabla_{\rho(\psi)} - {}^E\nabla_{\psi}\;.
\end{equation}
In the particular case of $V=E$ and
${}^E\nabla$~$=$~${}^E\Tilde{\nabla}$ it coincides with the
$E$-torsion tensor of $\nabla_{\rho(\cdot)}$, \cf~the text following
Eq.~\refeq{tildecon}. Thus $D\phi$ is indeed a section of $T^*\Sigma
\otimes \cX^*V$, as it should be.\footnote{For a more general ansatz
  of a covariant derivative in the Lie algebroid setting
  \cf~\cite{Melchior}, with however the same result.}
 
Now we can compute the gauge transformation of \refeq{eq:cov}. In this
context we will adopt a slightly vague, but, for practical purposes,
still quite useful point of view: We find that the covariant derivative
$D$ ``commutes'' with gauge transformations, but only modulo a term
proportional to the field strength $F^i$, 
\begin{equation}
  \label{eq:9} 
  (\delta_\epsilon D\phi)^\sigma - (D\delta_\epsilon\phi)^\sigma =
  \epsilon^aF^i\widetilde{S}_{ai}{}_\tau^\sigma\phi^\tau +
  \epsilon^aA^b\,{}^E\!R_{ba}{}_\tau^\sigma\phi^\tau\;.
\end{equation}
Indeed, the second term vanishes identically since ${}^E\nabla$ is a
flat $E$-connection. Here
$\widetilde{S}_{ai}{}_\tau^\sigma$~$\equiv$~$\big(\nabla_iT_a-\rho_a^jR_{ij}\big)_\tau^\sigma$.
We remark in parenthesis that for the adjoint $E$-connection
${}^E\Tilde{\nabla}$ on $E$ the tensor $S$ parametrizing the
non-closure of gauge transformations on $A^I$, \cf~Eq.~\refeq{var},
and $\widetilde{S}$ do, for $R_{ij}{}_c^a\neq 0$, not coincide,
$S_{ai}{}_b^c = \widetilde{S}_{ai}{}_b^c + \rho_a^jR_{ij}{}_b^c$.


The action of LAYM theory coupled to matter fields $\phi^\sigma$ is then
the sum of the LAYM action and a kinetic term for the mattter fields,
\begin{equation}
  \label{eq:LAYM_matter}
    S_{LAYM+matter} = S_{LAYM} -\int_\Sigma 
    \tfrac{1}{2}\, (D\phi)^\sigma\wedge\star (D\phi)^\tau \; g_{\sigma\tau}(X)\;,
\end{equation}
where $g_{\sigma\tau}\sim {}^V\!g$ is (the pullback by $\cX$ of) 
a non-degenerate metric on $V$. The action is
invariant under the gauge symmetries, if $g_{\sigma\tau}$ is compatible with
the $E$-connecton ${}^E\nabla$ on $V$, \ie~if 
\begin{equation} \label{covg}
  {}^E\nabla({}^V\!g)=0\,.
\end{equation}
The terms proportional to $\tilde S$, coming from the variation of the
kinetic term by use of eq.~\refeq{eq:9}, are proportional to $F^i$ and
thus can be absorbed by redefining $\delta_\epsilon B_i$
correspondingly.

It is easy to add \eg~a mass term for $\phi$ to this, using
${}^V\!g$: $\int_\Sigma \phi^\sigma \phi^\tau g_{\sigma\tau} \, \rm{vol}_\Sigma$ is
already by itself invariant under gauge transformations (here
$\rm{vol}_\Sigma$ denotes the volume form on $\Sigma$ induced by
$h$). This can be generalized in a straightforward manner to higher
powers in $\phi$, including thus self-interactions of the scalar
fields, by means of completely symmetric tensors $I_{\sigma_1\ldots \sigma_n}
\sim I \in \Gamma(S^n V)$  which are $E$-covariantly
constant, ${}^E\nabla(I)=0$: 
\begin{equation} \label{I} \sum_n
  \int_\Sigma \phi^{\sigma_1} \ldots \phi^{\sigma_n} I_{\sigma_1 \ldots \sigma_n}(X) \,
  \rm{vol}_\Sigma \;. 
\end{equation}


Another way of obtaining a coupling of a LAYM theory \refeq{eq:LAYM}
to scalar fields is to perform a Kaluza-Klein dimensional reduction
from $\Sigma_d$ to $\Sigma_{d-1}$ along a circle $S^1$, which we may
take to be along the direction $\mu=0$. Then the vector field $A^a$ on
$\Sigma_d$ decomposes into a vector field $\hat{A}^a$ on
$\Sigma_{d-1}$ and into a scalar field, $\phi^a$ coming from the
$0$-component of $A^a$. As shown in Appendix~B, the dimensional
reduction of both, the gauge symmetries and the action, shows that
$\phi^a$ transforms according to ${}^E\Tilde{\nabla}_a$,
Eq.~\refeq{tildecon}; moreover the $(0,m)$ component of $F^a$
coincides with the covariant derivative $D$ for the particular
$E$-covariant derivative $^E\nabla_I={}^E\Tilde{\nabla}_I$.  One might
ask how the condition of a flat $E$-connection found in this Section is
compatible with dimensional reduction where $^E\Tilde\nabla$ is
arbitrary. However, the dimensional reduction of the zero-component of
$F^i$ restricts $\phi^a$ to $\ker\rho$ on-shell---as a relict from the
$B_iF^i$-term in $S_{LAYM}$--- where the curvature of
$^E\Tilde\nabla$ vanishes, \cf~\refeq{ER=rhoS}. Dimensional reduction
therefore leads to a rather restricted setting. The $E$-connection is
permitted to be nonflat, but at the price of restricting the scalar
fields to taking values in $\ker\rho$ only.

\section{Matter fields of sigma model type}



One of the possible perspectives on a LAYM theory is that it
generalizes ordinary YM gauge theories to the realm of sigma models,
\cf,~\eg,~\cite{Srni}. In the usual YM setting, scalar
fields, like the Higgs field, take values in vector bundels associated
to the principal bundle in which the gauge fields are
connections. From the present perspective, such a restriction to
linearity, as present in the formulas \refeq{cov},
\refeq{eq:5}, \refeq{eq:6} for example, seems unnecessary and non-natural.  
In the ordinary Lie algebra situation,
$E$~$=$~$\Lieg$, this corresponded to linear representations of the Lie
algebra on a vector space $V$ (used in the construction of the
associated bundle). However, we may be interested also in
nonlinear, sigma-model like couplings of the scalar fields to the
LAYM-part. 

Towards this goal it is useful to note that the data used in the previous
section, a Lie algebroid $E \to M$ together with a flat $E$-connection
$^E\nabla$ on $p \colon V \to M$
can be put together into a bigger Lie algebroid $\tE$: As a vector
bundle this Lie algebroid is just $\tE\equiv p^*E \to V$, \ie~$E$
considered as living over the bundle $V$ as base manifold. The Lie
bracket between sections coming from sections of $E$ is the old one,
$[p^* \psi_1,p^* \psi_2] := p^* [\psi_1,\psi_2]$. It remains to define
what happens when $p^* \psi_2$ is multiplied by a function over $V$
that is fiber-linear (the rest follows by the Leibniz rule), \ie~by
sections $\alpha \in \Gamma(V^*)$. It is here where the $E$-connection
enters: $[p^* \psi_1,\alpha p^* \psi_2] := \alpha p^* [\psi_1,\psi_2]
+ ({}^E\nabla_{\psi_1}\alpha) \, p^* \psi_2$. The flatness condition
of $^E\nabla$ comes in when checking the Jacobi condition of that
bracket.

Now it is straightforward to generalize to the nonlinear setting. Let
us just replace the vector bundle $p \colon V \to M$ by a general
fiber bundle $p \colon \tM \to M$. Again we can consider the vector
bundle $\tE := p^* E \to \tM$.  Instead of a representation on $V$ we
want to consider the structure of a Lie algebroid defined on $\tE$,
satisfying an appropriate compatibility condition with $E$ certainly:
There is always a natural projection $\pi \colon p^* E \to E$ induced
by $p \colon \tM \to M$.
\begin{equation}
  \label{pullback-morphism}
  \begin{CD}
    \tE\equiv p^*E @>>>  \tM \\
   @V\pi VV            @VVpV \\
    E             @>>>  M
  \end{CD}
\end{equation}
One can check that in the linear situation above, $\pi$ is a Lie
algebroid morphism (\cf~\eg~\cite{BKS} for a convenient way of
checking this). This is what we now want to require also in the
present more general situation: by definition, an $E$-action on $\tM$
is a Lie algebra structure on $p^*E$ such that the projection $\pi$ is
a morphism of Lie algebroids.

It is important in this context that $\tE$ really \emph{is} the bundle
$p^*E$ and not just isomorphic to it and that $\pi$ is the
corresponding canonical projection.  One can check, furthermore, that
for $M$ being a point the Lie algebroid $\tE \to \tM$ reduces to the
action Lie algebroid $\tE = \Lieg \times \tM$ of a Lie algebra action
$\Lieg$ on a manifold $\tM$. So, $\tE = E \times_M \tM$ is the
``action Lie algebroid'' of a Lie algebroid $(E\to M)$-action on
$\tM\to M$.

Part of the Lie algebroid morphism property of $\pi \colon \tE \to E$
is the commutativity of the following diagram
\begin{equation}
  \label{connection}
   \qquad  \qquad \begin{CD}
    p^*E      @>\trho>> T\tM \\
   @V\pi VV           @VVp_*V \\           
    E        @>>\rho>  TM
  \end{CD}   \qquad     \qquad .
\end{equation}
This permits us to identify the anchor map $\trho \colon \tE\to T\tM$
with an $E$-connection on the fiber bundle $p \colon \tM\to M$, which,
by definition as given in \cite{Fernandes}, is precisely a map $\trho$
such that the above diagram is commutative. A map $\trho$ permits to
lift a vector $\psi_x \in E_x$ at the point $x\in M$ to the
corresponding vector in $T_u\tM$ at the point $u\in\tM$ with
$p(u)=x$. Commutivity of the diagram means that this ``horizontal
lift'' should be such that the projection down to $M$ by $p_*$ of the
lifted vector agrees with the vector $\rho(\psi_x)$. For $E=TM$, $\rho
= \id$, the standard Lie algebroid, this reproduces the standard
condition of an ordinary connection in $p \colon \tM \to M$ that the
composition of the projection with the lift is the identity on $T_xM$.

Since $\tE$ is a Lie algebroid, its anchor is a morphism of Lie
brackets, 
\begin{equation}  
  \trho\big([\tpsi_1,\tpsi_2]\big) 
  - 
  \big[\trho(\tpsi_1), \trho(\tpsi_2)\big]=0
\end{equation} 
which, for $\trho$ being viewed as an $E$-connection on $\tM$, is
tantamount to its flatness. In fact, a flat $E$-connection  $\trho
\colon p^* E \to T\tM$ on $\tM$ can be seen to be \emph{equivalent} to our
definition of an $E$-action on $\tM$. 
In this formulation we
easily reproduce the results of the previous section, where the
connection was further restricted to respect the linear structure on
the bundle $\tM = V$. 

We now put this into explicit formulas, generalizing the respective
ones of the previous section. In bundle coordinates $(X^i,\phi^\sigma)$ on
$\tM$, the anchor of $\tE$ applied to the ($\tM$-fiberwise constant)
basis $\te_a := p^* e_a$ induced by a local basis of sections on $E$,
takes the form
\begin{equation}
  \label{anchor}
  \trho(\te_a) = 
       \rho_a^i(x)             \frac{\partial}{\partial x^i} 
     + \trho_a{}^\sigma(x,\phi) \frac{\partial}{\partial \phi^\sigma}\;,
\end{equation}
where instead of $\trho_a{}^\sigma$ we could have written also
$\Gamma_a{}^\sigma$, stressing the interpretation of these components
as an $E$-connection on $\tM$. Equation (\ref{eq:5}) for the gauge
transformations now turns into
\begin{equation}
  \label{eq:5prime}
  \delta_\epsilon\phi^\sigma = -\epsilon^a\trho_a{}^\sigma(X,\phi)\;,
\end{equation}
while for the exterior covariant derivative of $\phi \in
C^\infty(\Sigma, \cX^*\tM)$ we get
\begin{equation}
  \label{eq:6prime}
  D\phi^\sigma = \rmd\phi^\sigma + \trho_a{}^\sigma (X,\phi) A^a +
  \Gamma_i{}^\sigma (X,\phi)  F^i\;.
\end{equation}
Here $\Gamma_i{}^\sigma$ denote the components of an ordinary connection on
$p \colon \tM \to M$. Requiring linearity in $\phi$, we
recover the context of the previous section  in all these formulas. 

Now we are in the position of considering the coupling of a kinetic
sigma model term to the pure gauge part of the action. This gives
\begin{equation}
  \label{eq:LAYM_mattersigma}
  S_{LAYM+matter} = S_{LAYM} -\int_\Sigma 
  \tfrac{1}{2}\, (D\phi)^\sigma\wedge\star (D\phi)^\tau \; g_{\sigma\tau}(X,\phi)\;.
\end{equation}
The allegedly small change of permitting $g_{\sigma\tau}$ to depend
also on $\phi$ in comparison to (\ref{eq:LAYM_matter}) implies some
conceptual complications: Before, $g_{\sigma\tau}$ corresponded to a
fiber metric on $V$, which we could also view as a quadratic
\emph{function} on $V=\tM$. Now, $^Vg$ is a fiber metric on $V\tM
\subset T\tM$, the subbundle \emph{over} $\tM$ consisting of vertical
tangent vectors. A condition of the type (\ref{covg}) does not yet
make any sense thus, we first need a Lie algebroid-connection on
$V\tM$, which can be viewed also as the foliation Lie algebroid
$T{\cal F}$ of the foliation/fibration of $\tM$ by its fibers.

However, in fact there is a \emph{canonical} lift of the flat 
$E$-connection $\trho \colon \tE \to T\tM$ to a flat $\tE$-connection
$\ttrho \colon \ttE \to T(V\tM)$ with $\ttE = \tp^* \tE$ and $\tp
\colon V \tM \to \tM$:
\begin{equation}
  \label{connection2}
   \qquad  \qquad \
   \begin{CD}
    \ttE\equiv\tp^*\tE  @>\ttrho>> T(V\tM) \\
   @V\tpi VV            @VV\tp_*V \\           
    \tE  \equiv p^*E     @>\trho>>  T\tM \\
   @V\pi VV             @VVp_*V \\           
    E         @>>\rho>   TM
  \end{CD}   
  \qquad     \qquad .
\end{equation}
In other words, there exists a Lie algebroid 
structure on $\ttE = \tp^* \tE$ such that $\tpi \colon \ttE \to \tE$
is a Lie algebroid morphism: 
\begin{equation}
  \label{connection3}
   \qquad  \qquad \
   \begin{CD}
    \tp^*\tE  @>>> V\tM \\
   @V\tpi VV            @VV\tp V \\           
    \tE       @>>>  \tM \\
  \end{CD}   
  \qquad     \qquad .
\end{equation}
In order to show that the Lie algebroid structure on $\tE$ induces a
Lie algebroid structure on $\ttE$, it suffices to specify the anchor
$\ttrho$ of the latter, since the bracket on $\ttE$ is fixed
already uniquely by means of the bracket on $\tE$ or $E$ when applied
to sections coming from $\tE$ and $E$, respectively. Denoting by
$\varphi^\sigma = \rmd \phi^\sigma$ fiber linear coordinates on $V\tM$, the
anchor map $\ttrho$ of $\ttE$ applied to $\tte_a : = \tp^* \te_a
\equiv \tp^* p^* e_a$ reads as
\begin{equation}
  \label{eq:ttrho}
  \ttrho_a =  \rho_a^i(x)         \frac{\partial}{\partial x^i} 
  + \trho_a{}^\sigma(x,\phi) \frac{\partial}{\partial \phi^\sigma} 
  +  \frac{\partial\trho^\sigma_a(x,\phi)}{\partial\phi^\tau}
  \varphi^\tau
  \frac{\partial}{\partial\varphi^\sigma} \;.
\end{equation}
This also corresponds to a flat $\tE$-connection on $V\tM$ with
components
$\tGamma_a{}^\sigma_\tau(x,\phi)$ $=$ $\frac{\partial}{\partial\phi^\tau}\trho^\sigma_a(x,\phi)$.

Let us now provide a coordinate independent construction of this
canonical lift, which also shows that its definition is independent of
the chosen basis in $E$, and that the construction depends crucially on
restriction to vertical vector fields on $\tM$ (equipped itself with a
flat $E$-connection). We want to define a bundle map $\ttrho \colon
\ttE \to T(V\tM)$. Extend a point $\ttpsi_0 \in \ttE$ to some
fiberwisely constant section $\ttpsi \in \Gamma(\ttE)$ coming from a
section $\psi\in \Gamma(E)$; so, in the previously introduced local
basis of sections in $\ttE$, $\ttpsi = \psi^a \tte_a$ with $\psi^a$
depending on coordinates $x^i$ of $M$ only and with $\ttpsi$ evaluated
at the projection of $\ttpsi_0$ to $M$ agreeing with $\ttpsi_0$. This
induces also a section $\tpsi = \psi^a \te_a$ in $\tE$, whose image
with respect to $\trho$ gives a vector field on $\tM$. Consider the
(local) flow $\Phi^t_\psi$ of this vector field and lift it to $T\tM$
by means of the pushforward map $(\Phi^t_\psi)_* \colon T\tM \to
T\tM$, a vector bundle morphism covering the flow $\Phi^t_\psi$ on
$\tM$. This lift is thus generated by a vector field on $T\tM$
covering the vector field $\trho(\tpsi)$. We can restrict the vector
field viewed as a section in $T(T\tM)$ to the submanifold $V\tM$ of
$T\tM$. Two things happen in this context: Firstly, while the vector
field on $T\tM$ is not $\mathcal{C}^\infty(M)$ linear in $\psi \in
\Gamma(E)$ in general, the restriction has this property (which is
essential for having the result being independent on the extension of
$\ttpsi_0$ to an at least locally defined section $\ttpsi$ or
$\psi$). Secondly, the restriction is tangent to $V\tM \subset T\tM$
(here the fact that $\trho(\tpsi)$ is projectable to $M$, covering
$\rho(\psi)$, \cf~diagram (\ref{connection}), enters crucially) and
can thus be viewed as a vector field on $V\tM$. Evaluate this vector
field at the point in $V\tM$ living under $\ttpsi_0 \in \ttE$ and call
this $\ttrho(\ttpsi_0)$. By a straightforward calculation one may
check that this geometric construction indeed yields (\ref{eq:ttrho}).

With these ingredients at hand, we are now in the position to
formulate a condition on the fiber metric $^Vg$ on $V\tM$ as entering
the functional (\ref{eq:LAYM_mattersigma}). The functional becomes
invariant w.r.t.~gauge transformations if the following condition on
$g$ is satisfied (in addition to the conditions to be placed on
${}^E\!g)$:
\begin{equation} \label{covgsigma}
{}^{\tE}\nabla(g)=0 \, ,
\end{equation}
where ${}^{\tE}\nabla$ is the flat $\tE$-connection corresponding to
(\ref{eq:ttrho}) and described in the sentence after that formula.

In the present more general framework than in the previous section,
formulating the conditions on some selfinteraction for the scalar
fields, \cf~Eq.~\refeq{I} and the corresponding discussion, becomes
simpler: We can add to \refeq{eq:LAYM_mattersigma} any term of the
form
\begin{equation}
  \int_\Sigma W(X,\phi) \rm{vol}_\Sigma \, ,
\end{equation}
provided only that $W$ is a function on $\tM$ invariant along the
orbits generated by $\trho$, \ie~if $(\rho_a^i\partial_i+\trho_a^\sigma\partial_\sigma)W=0$.

If the Lie algebroid $E$ permits an integration to an source-simply
connected Lie groupoid $\G\rightrightarrows M$
(\cf~\cite{CF,LecturesCrainicFernandes} for the necessary and
sufficient conditions), the above considerations have the following
global reinterpretation: First, given $\G$ we can consider its action
on $p \colon \tM \to M$, where $p$ is usually called the moment map in
this context. An action is then given by a map $\varphi \colon \G
\times_M \tM \to \tM$ which is compatible with the structural maps on
$\G$. In particular this means that any $g\in
\G$ with source $x$ and target $y$ is lifted to an isomorphism of
fibers, $\varphi_g \colon p^{-1}(x) \to p^{-1}(y)$. This can again be
made into a new groupoid $\widetilde{\G} \rightrightarrows\tM$ whose
elements $\widetilde{g}$ consist of the maps $\varphi_g$ mapping one
point in $\tM$ (the source of $\widetilde{g}$) to another one (the
target of $\widetilde{g}$). Finally, the diffeomorhpisms of
$\tM$-fibers $\varphi_g$ can be lifted to isomorphisms of their
tangent bundles. This induces canonically a Lie groupoid
$\widetilde{\widetilde{{\G}}} \rightrightarrows T{\cal{F}}$. As
already anticipated by the notations, these two groupoids are the
integrations of $\tE$ and $\ttE$, respectively, as we recommend the
reader to check as an exercise. 
The condition (\ref{covgsigma}) now just states
that the maps $(\varphi_g)_* \colon T(p^{-1}(x)) \to T(p^{-1}(y))$,
corresponding to a collection of elements in
$\widetilde{\widetilde{{\G}}}$, are also isomorphisms (isometries) of
$T{\cal{F}}$ equipped with the fiber metric $^Vg$. 

\section{Continuative Discussion}
\label{Discussion}
In this concluding section we want to discuss two more aspects of the
topics presented in this article up to here. First of all this
concerns the pure gauge field system (\ref{eq:LAYM}), relaxing the
conditions on the tensor ${}^Eg$ needed for squaring the 2-form field
strength. Afterwards we come back to the coupled matter gauge field
system, discussing it from a slightly more unified perspective. We now
turn to the first issue.

The field strengths $F^i$, $F^a$ satsify some generalized version of
Bianchi identities \cite{Strobl}.\footnote{In fact, such an
  observation may be even used as a starting point for constructing
  algebroid type gauge theories, containing also nonabelian gerbes, as
  demonstrated in detail in \cite{Melchior}.} In what follows in
particular the first one of those will play an important role, for
which reason we display it explicitely: 
\begin{equation} \label{Bianchi1} 
  \rmd F^i - \rho^i_a,_j A^a \wedge F^j + \rho^i_aF^a
  =0 \, . 
\end{equation}

Primarily, this leads to a second independent gauge symmetry
\cite{Strobl}. Suppose that we transform $B_i$ according to
\begin{equation} \label{lambda}
  \delta_\lambda B_i := \rmd \lambda_i + \rho^j_a,_i A^a \wedge \lambda_j\,.
\end{equation}
Then it is easy to see that $S_{LAYM}$ is invariant w.r.t.~such
transformations up to boundary contributions (resulting from a partial
integration), if $\lambda_i \rho^i_a =0$---implying, more
geometrically, that $\lambda$, instead of taking values arbitrarily in
$T^*M$, is restricted to the conormal bundle of the tangent
distribution to the orbits generated by $\rho$. In fact, here, and
also in what is to follow, we will consider only regions of $M$ where
the rank of $\rho$ is constant. Further investigations of what happens
more precisely at regions where the rank of $\rho$ jumps would be
interesting though.

One may employ Eq.~\refeq{Bianchi1} in another direction also,
however: The contraction of $\rho$ with the 2-form field strength can
be expressed in terms proportional to the 1-form field strengths (and
its derivative). Since, on the other hand, any term proportional to
$F^i$ in \refeq{eq:LAYM} can be dropped by an appropriate redefinition
of the field $B_i$, one finds that there is an equivalence relation
between fiber metrics on $E$ yielding physically equivalent gauge
theories---in fact, the tensors ${}^Eg$ can even become partially
degenerate by such redefintions. Let $e^a$ denote a local frame in
$E^*$, then ${}^Eg = g_{ab} e^a e^b$. Consider replacing $g_{ab}$ by
$\bar{g}_{ab} = g_{ab} + \rho_a^i \beta_{ib} + \rho_b^i \beta_{ia}$
for some collection $\beta_a$ of 1-forms on $M$.  Since in the action
functional ${}^Eg$ is contracted with $F^a$s, the terms proportional
to $\beta_{ai}$ can be absorbed completely: we replace $\rho(F_{(2)})$
by the corresponding two terms according to \refeq{Bianchi1}, perform
a partial integration in the term with $\rmd F^i$, and then absorb all
prefactors of the newly introduced terms proportional to $F^i$ by
redefining $B_i$ appropriately. This means that a redefinition 
$g_{ab} \mapsto \bar{g}_{ab}$ can be compensated by a local
diffeomorphism on the field space of the functional \refeq{eq:LAYM}. 
In other words, on the physical level, there is an equivalence between
of two functionals \refeq{Bianchi1} induced by an equivalence relation
between its $E$-2-tensors ${}^Eg$
\begin{equation}
  g_{ab} \sim g_{ab} + \rho_a^i \beta_{ib} + \rho_b^i \beta_{ia} \,  
\end{equation}
for arbitrary choices of $\beta \in \Omega^1(M,E^*)$. The quotient of
$\Gamma(M,S^2 E^*) \ni {}^Eg$ by these orbits is in one-to-one
correspondence to fiber metrics on the subbundle $\ker \rho
\subset E$.\footnote{This is true, when restricting to orbits that 
have at least one non-degenerate representative ${}^Eg$. Recall also
that $E$ was assumed to be regular for the moment so that under this
assumption its kernel really defines a subbundle of $E$.}  
Denote the \emph{restriction} of ${}^Eg$ to $\ker\rho$ by ${}^\rho
g$; it is one-one to some equivalence class $[{}^Eg]$ of a fiber 
metric on $E$. 

The bundle $\ker \rho \to M$ carries a \emph{canonical}
$E$-connection. Let $\psi \in \Gamma(E)$ and $\tilde{\psi} \in
\Gamma(\ker \rho)$ and define
\begin{equation} \label{Bott} 
  {}^\rho{\nabla}_\psi \Tilde{\psi} :=  [\psi, \Tilde{\psi}] \, .
\end{equation}
Since $\tilde{\psi}$ is in the kernel of $\rho$, this is indeed
$C^\infty(M)$-linear in $\psi$. This connection is sometimes called
the $E$-Bott connection. Comparing with equations (\ref{condition})
and (\ref{tildecon}), it is now obvious that 
\begin{equation} \label{newcondition} 
  {}^\rho{\nabla}\; {}^\rho g = 0 
\end{equation} 
is sufficient for gauge invariance of \refeq{eq:LAYM}. In contrast to
\refeq{condition}, this condition is not only independent of any
auxiliary connection $\nabla$ on $E$, it is certainly also a weaker
condition on ${}^Eg$, needing ${}^Eg$ only to be in some orbit
characterized by its restriction ${}^\rho g$ to $\ker \rho$ such that
\refeq{newcondition} holds true. 

\vskip5mm

We now turn to the second issue, the coupled matter gauge field
system. The emphasis on a new Lie algebroid $\tE \to \tM$ governing
linear or nonlinear actions of Lie algebroids $E\to M$ on bundles $\tM
\to M$ corresponding to matter field target spaces also resides in a
possible reinterpretation of the gauge invariant coupled matter-LAYM
functional (\ref{eq:LAYM_mattersigma}). Who forbids one to consider
all the coordinates on $\tM$ on the same footing to start with. We had
the kind of no-go theorem around (\ref{Fhoch2}), where a squaring of
all 1-form and 2-form field strengths was
taken. Eq.~(\ref{eq:LAYM_mattersigma}) from this perspective shows
that squaring some of the 1-form field strengths, keeping the others
included via Lagrange multipliers, does not necessarily lead to
likewisely strong restrictions on admissible Lie algebroids. At the
same time, some of the coordinates of the target Lie algebroid are
promoted into propagating degrees of freedom typical for scalar fields
from the physical point of view.

We make this point more explicit by rewriting
(\ref{eq:LAYM_mattersigma}) in this spirit. First, we denote by
$x^I=(x^i,\phi^\sigma)$ collectively all coordinates on $\tM$ and,
correspondingly, by $\widetilde{\cX}$ the map from $\Sigma$ to all of
$\tM$.
Then, the corresponding 1-form field strengths $F_{(1)}^I$ split into
$F_{(1)}^i$, agreeing with the respective previous expression
(\ref{eq:f}) except that, for clarity, we added an index in brackets
to emphasize to form character of the field strength, and, by the same
formula, one now has $F_{(1)}^\s = \rmd \phi^\s - \trho^\s_a
A^a$. Note that geometrically $F_{(1)}^i$ corresponds to elements
tangent to $M$ and $F_{(1)}^\s$ tangent to fibers of $\tM \to
M$. $F_{(1)}$ should be an element of
$\Omega^1(\Sigma,\widetilde{\cX}^* T\tM)$ on the other hand, \ie~a
vector on $\tM$. The two components cannot be combined intrinsically
or coordinate independently into a meaningful vector on $\tM$ without
a connection on that bundle. Let $(\rmd \phi^\s + \Gamma_i^\s \rmd
x^i) \frac{\partial}{\partial \phi^\s} \in \Omega^1(\tM, V\tM)$ be
such a connection 1-form on $\tM$,
its kernel determining what is horizontal in $T\tM$, 
\begin{equation}\label{eq:split}
V\tM \oplus H\tM = T\tM \quad \ni \quad v = v^{ver} + v^{hor} \; .
\end{equation}
We now see that the vertical part of $F_{(1)}$, $F_{(1)}^{ver}=
F_{(1)}^\s + \Gamma_i^\s F_{(1)}^i$, reproduces precisely
eq.~(\ref{eq:6prime}). On the other hand, the horizontal part is
always proportional to $F_{(1)}^i$ (for any choice of $\Gamma_i^\s$),
thus the first term in (\ref{eq:LAYM}) constrains $F_{(1)}^{hor}$ to
vanish. Likewisely, we could map $F_{(1)} \in
\Omega^1(\Sigma,\widetilde{\cX}^* T\tM)$ by $p_* \circ \widetilde{\cX}$ to
a tangent component on the base $M$ of $\tM$ and interpret the first
LAYM-term in this way within the present setting, the Lagrange
multiplier living in $T^*M$ then as before. Preferring the first
option, a coordinate independent, geometrical form of the total
action, using $\tE \to \tM$ as starting Lie algebroid and a split of
$T\tM$ into the two subbundles as above in (\ref{eq:split}), one finds
for the combined matter-gauge field action (\ref{eq:LAYM_mattersigma})
the following form:
\begin{equation} \label{tildeaction} 
  \int_\Sigma \langle B
  \stackrel{\wedge}{,} F_{(1)}^{hor} \rangle - \tfrac{1}{2} 
  \left({}^{V\tM}\!g \circ \widetilde{\cX}\right) 
  \left( F_{(1)}^{ver} \stackrel{\wedge}{,} \star F_{(1)}^{ver} \right)
  -
  \tfrac{1}{2}\, \left( {}^{\tE}g \!\circ \widetilde{\cX}\right) 
  \left( F_{(2)}\stackrel{\wedge}{,} \star F_{(2)} \right)\;,
\end{equation}
where $B$ is a $(d-1)$-form taking now values in (the pullback by
$\widetilde{\cX}$ of) $H^*\tM$.

This shows that partially squaring some of the 1-form field strengths
is compatible with Lie algebroids different from mere Lie
algebras. Gauge invariance of such a functional will certainly also
heavily restrain the starting Lie algebroid $\tE \to \tM$. What we
showed constructively is that such a functional is compatible with a
Lie algebroid structure on $\tE$ coming from a Lie algebroid $(E \!
\to \!  M)$-action on $p \colon \tM \! \to \! M$ for some $E$ and $M$
such that one has the diagram (\ref{pullback-morphism}) with $\pi$
being a Lie algebroid morphism. In the language of \cite{KotovStrobl}
this corresponds to a Q-bundle $\pi\colon \tE[1] \to E[1]$ (which is
locally trivial only in the sense of graded but not in the category of
Q-manifolds), where in any local chart on the total space there exists
a canonical isomorphism of its degree one veriables with the degree
one variables on the base.

In the extreme context of squaring \emph{all} 1-form field strengths
we showed that one is \emph{necessarily} lead to the Lie algebroid of
a Lie algebra action on its base. This corresponds to the situation of
$M$ being a point in the discussion above. It may be interesting to
see if in a generalization of this observation a functional of the
form (\ref{tildeaction}) with Lie algebroid $\tE$ always leads to the
scenario as in (\ref{pullback-morphism}) above.


\appendix
\section{Some formulas on Lie algebroids}

A Lie Algebroid consists of a vector bundle $E\rightarrow M$ over a
manifold $M$, a Lie algebra bracket, $[\cdot, \cdot]$, between
sections $\psi$ of $E$, and of a bundle map $\rho: E\rightarrow TM$,
called the anchor map. The bracket satisies a Leibnitz rule,
\begin{equation}
  [\psi_1,f\psi_2] = f[\psi_1,\psi_2] + \rho_{\psi_1}(f)\psi_2\,,
  \quad f\in\mathcal{C}^\infty\;,\quad \psi_1,\psi_2\in\Gamma(E)\,.
\end{equation}
In local coordinates $X^i$ on $M$ and a local basis $e_a$ of $\Gamma(E)$,
this data is encoded in structural functions
$C_{ab}^c,\rho_a^i\in\mathcal{C}^\infty(M)$, such that the bracket and
the anchor map take the form $[e_a,e_b]$~$=$~$C_{ab}^c\,e_c$, and
$\rho(e_a)$~$=$~$\rho_a^i\partial_i$.
As a consequence of the definitions above, the anchor map is a
morphism wrt.~the bracket, \ie
\begin{equation}
  [\rho(e_a), \rho(e_b)] = \rho([e_a,e_b])\,.
\end{equation}
Examples of Lie Algebroids include a bundle Lie
Algebras ($\rho=0$), $TM$~($\rho=id$), and Poisson manifolds.

In order to talk about $E$-connections ${}^E\nabla$, we need to
specify the Leibnitz rule:
\begin{equation}
  \label{eq:14}
  {}^E\nabla_{\psi_1}(f\psi_2) = 
  f\;{}^E\nabla_{\psi_1} \psi_2 + \rho_{\psi_1}(f) \psi_2\;,
\end{equation}
Any connection $\nabla$ on the vector bundle $E$ can be lifted to an
$E$-connection using the anchor map: $\nabla_{\rho(\cdot)}$. 

Now that we have the concept of an $E$-connection on a Lie Algebroid,
we can translate concepts involving connections on vector bundles to
the realm of Lie Algebroids. For a connection $\nabla$ on a vector
bundle, the curvature is defined as
\begin{equation}
  \label{eq:curv}
  R(\partial_i, \partial_j) = 
  \nabla_i\nabla_j - \nabla_j\nabla_i - \nabla_{[\partial_i,\partial_j]}\;.
\end{equation}
Analogously, we define the corresponding $E$-curvature as
\begin{equation}
  \label{eq:e-curv}
  {}^E\!R(\psi_1,\psi_2) = 
  {}^E\nabla_{\psi_1}{}^E\nabla_{\psi_2} - 
  {}^E\nabla_{\psi_2}{}^E\nabla_{\psi_1} - 
  {}^E\nabla_{[\psi_1,\psi_2]}\;.
\end{equation}
By the morphism property of the anchor map the $E$-curvature of an
induced $E$-connection satisfies
\begin{equation}
  {}^E\!R(\psi_1,\psi_2) = R\big(\rho(\psi_1),\rho(\psi_2)\big)\,.
\end{equation}

Given any $E$-connection $^E\nabla$, we can form a \emph{tensor}
involving the structure functions $C_{ab}^c$. This tensor
$T\in\Omega^1(E)\otimes\Gamma(End(E))$ is called the the $E$-torsion
tensor corresponding to the $E-$connection $^E\nabla$ and is defined
as
\begin{equation}
  \label{eq:e-tors}
  T(\psi_1) \psi_2 = 
  [\psi_2,\psi_1] + {}^E\nabla_{\psi_1} \psi_2 - {}^E\nabla_{\psi_2} \psi_1\;,
\end{equation}
which in components takes the form
\begin{equation}
  \label{eq:T}
  T(e_a)e^c = T_{ab}^ce^b\;,\quad
  T_{ab}^c = -C_{ab}^c + \Gamma_a{}_b^c - \Gamma_b{}_a^c\;.
\end{equation}

Finally, we derive an identity which involving the $E$-torsion of an
induced $E$-connection. In components, the induced connection is given
by $\Gamma_a{}_b^c$~$=$~$\rho_a^i\Gamma_i{}_b^c$.  As a consequence of
the Jacobi identity, the $E$-torsion corresponding to this induced
connection satisfies the identity:
\begin{equation}
  \label{eq:12}
  T_{ab}^dT_{cd}^e + cycl(abc) = \rho_c^i\nabla_iT_{ab}^e 
  + \rho_c^i\rho_b^jR_{ij}{}_a^e + cycl(abc)\;,
\end{equation}
which can be used to show that $E$-curvature of the ``adjoint connection''
${}^E\Tilde{\nabla}_a$ (\cf~Eq.~\refeq{eq:4b}) reduces to
\begin{equation}
  \label{eq:S}
  {}^E\Tilde{R}_{ab}{}_c^d = \rho_b^iS_i{}_{ab}^d\;,\qquad
  S_i{}_{ab}^d = \nabla_iT_{ab}^d + \rho_b^jR_{ij}{}_a^d - \rho_a^jR_{ij}{}_b^d\;.
\end{equation}

\section{Dimensional Reduction of LAYM}

In ordinary YM theory scalar fields in the adjoint representation can
be obtained my performing a Kaluza-Klein dimensional reduction along a
circle $S^1$. Here, we perform this dimensional reduction for the LAYM
theory \refeq{eq:gt1}, \refeq{eq:gt2}, \refeq{eq:LAYM}.  Starting with
a $d$-dimensional world sheet $\Sigma_d$ we perform a dimensional
reduction to a $(d-1)$-dimensional world sheet $\Sigma_{d-1}$ by
splitting $\Sigma_{d}= \Sigma_{d-1}\times S^1$ and shrinking the
radius of the circle $S^1$ to zero. Then the components of $A$ along
the $S^1$-direction become scalar fields in the lower-dimensional
theory.

On $\Sigma_d$ we decompose the world-sheet indices $\mu=0...(d-1)$
into $(\mu)=(0,m)$ where $\mu=0$ denotes the direction along the $S^1$
and $m=1...(d-1)$ the directions perpendicular to that. The 1-form
fields $A^a$ split into$(A^a_\mu) = (A_0^a,A_m^a)$. After the
dimensional reduction, the zero components of $A_0^a$ become scalar
fields $\phi^a$ on $\Sigma_{d-1}$. The zero-component of the gauge
transformation of $A^I$ reduces to the gauge variations of $\phi^a$:
\begin{equation}
  \delta A_0^a \rightarrow \delta\phi^a 
  = c^a_{bc}\phi^b\epsilon^c - \Gamma_i{}_b^a\epsilon^b\rho_c^i\phi^c
  = -\epsilon^b\Tilde{\Gamma}_{bc}^a\phi^c\;,
\end{equation}
the reduction of the zero component of the $F^i=0$ field equations, $F^i_0
\rightarrow -\rho_a^i\phi^a$, constrains $\phi^a$ to be in $\ker\rho$,
and the reduction of the $(0,m)$ component of the field strength $F^a$
becomes the covariant derivative for $\phi^a$,
\begin{equation}
  F^a_{0,m}\rmd u^m \rightarrow \tfrac{1}{2} (D\phi)^a\;.
\end{equation}
Hence, the gauge transformations and the covariant derivative of
$\phi^a$ obtained by dimensional reduction coincide with the gauge
transformations and covariant derivative of a scalar field which takes
values in $E$ and transforms according to the adjoint connection
$^E\Tilde{\nabla}$. The difference between the two constructions is
that the fields generated by dimensional reduction are always in the
adjoint representation, and that they are constrained to taking values
in $\ker\rho$.

\end{document}